\documentclass[journal]{IEEEtran}
\usepackage[utf8]{inputenc}
 \usepackage{multirow}
 \usepackage{fixltx2e}
 \usepackage{caption, subcaption}
\usepackage{floatrow}
\usepackage{graphicx}%
\usepackage{dblfloatfix }
\restylefloat{table}
\usepackage[T1]{fontenc}
\usepackage{babel}
\usepackage[font=small,labelfont=bf,tableposition=top]{caption}
\usepackage{blindtext}
\usepackage{cuted}
\newtheorem{theorem}{Theorem}[section]
\newtheorem{defn}[theorem]{Definition}
\title{A Natural Language Processing Approach for Instruction Set Architecture Identification}
\usepackage{geometry}
\renewcommand{\thetable}{\arabic{table}}

\author{Dinuka Sahabandu$^{1}$,\IEEEmembership{~Student Member, IEEE,} Sukarno Mertoguno$^{2}$,\IEEEmembership{~Senior Member, IEEE,} and \\ Radha Poovendran$^{1}$,\IEEEmembership{~Fellow, IEEE}
\thanks{$^{1}$Dinuka Sahabandu and Radha Poovendran are with the Network Security Lab, Department of Electrical and Computer Engineering,
        University of Washington, Seattle, WA 98195-2500.
        {\tt\small \{sdinuka,rp3\}@uw.edu}}%
\thanks{$^{2}$Sukarno Mertoguno is with the  Institute for Information Security \& Privacy, Georgia Institute of Technology,
North Avenue,
Atlanta, GA 30332.
        {\tt\small sukarno.mertoguno@gtri.gatech.edu}}
}

\begin{document}

\maketitle

\begin{abstract}

 Binary analysis of software is a critical step in cyber forensics applications such as program vulnerability assessment and malware detection. This involves interpreting instructions executed by software and often necessitates converting the software's binary file data to assembly language.
{\color{black} The conversion process requires information about the binary file's target instruction set architecture (ISA).} However, ISA information might not be included in binary files due to compilation errors, partial downloads, or adversarial corruption of file metadata. 
Machine learning (ML) is a promising methodology that can be used to identify the target ISA using binary data in the object code section of binary files. 
{\color{black} In this paper we propose a binary code feature extraction model to improve the accuracy and scalability of ML-based ISA identification methods.  
Our feature extraction model can be used in the absence of domain knowledge about the ISAs.} Specifically, we adapt models from natural language processing (NLP) to i)~identify successive byte patterns commonly observed in binary codes, {\color{black} ii)~estimate the significance of each byte pattern to a binary file, and iii)~estimate the relevance of each byte pattern in distinguishing between ISAs.} We introduce character-level features of encoded binaries to identify fine-grained bit patterns inherent to each ISA.  We use a dataset with binaries from 12 different ISAs to evaluate our approach. {\color{black} Empirical evaluations show that using our byte-level features in ML-based ISA identification results in an $8\%$ higher accuracy than the state-of-the-art features based on  byte-histograms and byte pattern signatures. We observe that character-level features allow reducing the size of the feature set by up to 16x while maintaining accuracy above $97\%$.}

\end{abstract}

\section{Introduction} 


\textcolor{black}{
Modern technological devices such as  computers and mobile phones contain firmware and software that consist of thousands of lines of source code \cite{dit2013feature}.}
Developers deliberately make the source code unavailable to the public for proprietary and security reasons \cite{rajba2021information, henry1988technique}.
Methods to restrict access to the source code include encryption and obfuscation in order to restrict software piracy and mitigate malicious code injection \cite{behera2015different,naumovich2003preventing}. On the other hand, the source code of malicious software (malware) can also be obfuscated by cyber adversaries to avoid being analyzed by security experts \cite{moser2007limits}. Consequently, cyber forensics applications such as assessing potential program vulnerabilities \cite{sun2020hybrid,lee2013function} and malware detection \cite{han2013malware,willems2007toward,canfora2014metamorphic} study the functionality of software by analyzing their binary files. Binary files store the compiled source code as a string of ``0"s and ``1"s (binaries) interpretable to computer processors. Identifying control paths, data flows, and data types required for assessing content and structure of undisclosed source code often requires converting binaries to assembly language  \cite{andriesse2018practical,kapoor2004approach, xue2019machine}. This is called {\em disassembly process}.

The disassembly process requires information about the type of processor Instruction Set Architecture (ISA) on which the binaries are expected to run, instruction length, and endianness \cite{andriesse2018practical} (Examples in Table~\ref{tab:Disassembly-info}).
\setlength\tabcolsep{14pt} 
\begin{table}[H]
\begin{tabular}{|l|l|}
\cline{1-2}
\multicolumn{1}{|c|}{\textbf{Disassembly Info.}}                        & \multicolumn{1}{c|}{\textbf{Examples}}                                             \\ \cline{1-2}
\begin{tabular}[c]{@{}l@{}}Instruction set \\ architecture\end{tabular} & \begin{tabular}[c]{@{}l@{}}ARM, MIPS, x86, AVR, \\ PowerPC, SPARC, s390\end{tabular}  \\ \cline{1-2}
Instruction length                                                      & 16-bit, 32-bit, 64-bit                                                            \\ \cline{1-2}
Endianness                                                              & little-endian, big-endian                                                          \\ \cline{1-2}
\end{tabular}
\caption{Examples for different disassembly information}\label{tab:Disassembly-info}
\end{table}
\noindent {\color{black} The disassembly information is extracted from the file header (e.g., ELF header, PE header) or file name (e.g., .exe for 32-bit x86, 64.exe for 64-bit x86) of binary files. } However, these meta-data can be incomplete or missing due to compilation errors or partial downloads. Cyber adversaries have been observed to tamper with these  metadata fields in order to remain hidden after carrying out an attack \cite{cozzi2018understanding, mcintosh2021enforcing}.



In the absence of information on ISA type, multi-architecture disassemblers such as Ghidra~\cite{GHIDRA}, IDA Pro~\cite{IDA-pro}, and Capstone~\cite{capstone_2020}, and firmware analysis tools such as Binwalk~\cite{refirmlabs} have been employed to predict the target ISA. Such methods are based on brute forced disassembling of binaries and inspecting assembly codes for known architecture-specific signatures.
\textcolor{black}{However, modern computer systems like severs consist of multiple devices such as CPUs, GPUs, and network cards, each of which uses different types of virtual and physical process architectures~\cite{hennessy2011computer}.} 
\textcolor{black}{ The increasing number of different ISA types makes brute forced disassembling-based ISA identification computationally infeasible. The lack of information on target ISA can thwart the disassembly process and affect many cyber forensics applications such as software security assessments \cite{liu2013binary, khadra2016speculative} and cyber threat hunting \cite{yazdinejad2020cryptocurrency,  rad2012opcodes,bilar2007opcodes}.} Consequently, methods that enable accurate prediction of target ISA solely based on information extracted from (partial) binaries have gained attention recently.


Machine Learning (ML) has been explored as a methodology to predict ISA using features extracted from the object code section of binary files (e.g., byte-histograms and byte signatures) \cite{sickendick2013file, ma2019svm, clemens2015automatic,beckman2020binary, nicolao2018elisa, kairajarvi2020isadetect}. However,  accuracy and scalability of ML-based techniques are affected by the following limitations of  existing object code features. \textcolor{black}{ (i)~{\bf Lack of generalizability}: Preselected byte signatures included in feature set for capturing properties of ISA such as endianness may not necessarily be present in some (partial) binaries,  (ii)~{\bf Noisy data}:  The byte patterns that are commonly observed across the binaries of different ISAs  (e.g., 0x00\footnote{This paper used the standard prefix notation 0x to indicate that the subsequent number is in the hexadecimal format.} byte patterns) are assigned high importance, leading to erroneous predictions of ISAs, and (iii)~{ \bf Low resolution}: Byte-level granularities might not capture fine-grained bit patterns embedded in the object code of binary files.}
Enhancing the capabilities of ML-based ISA identification therefore requires an approach to characterize features that will highlight ISA-specific bit patterns that are frequently present in binaries of the corresponding architecture. 

\textcolor{black}{
We observe that natural language processing (NLP) research provides wide range of models to extract features from text data for applications such as document classification and email filters \cite{chowdhary2020natural}. We then observe that the object code binaries (i.e., machine language) which defines instructions for machines are generated according to a set of well-defined rules similar to the natural language texts that convey messages to human. Based on these two observations, we propose to use models from natural language processing (NLP) to extract features from  object code binaries.  Specifically, we use the N-gram term frequency-inverse document frequency (TF-IDF), a widely explored technique in NLP research for extracting a set of informative and discriminating text patterns \cite{ramos2003using, aizawa2003information}.
{\color{black} The binary code feature extraction techniques that we introduce do not require any expert knowledge about the ISAs.}
}
We make the following contributions.
\begin{itemize}
    \item We observe that successive bytes in object code binaries have correlated patterns specific to the ISA. We use byte-level $N$-gram TF features to extract such patterns.
    \item We scale byte-level $N$-gram TF features using their respective IDF values to reduce the effect of noisy data and increase the sensitivity to byte patterns that characterize the ISA.
    \item We observe that instruction bytes of binaries have fine grained bit patterns specific to the ISA and we use character-level $N$-gram TF-IDF features of encoded binaries to extract such patterns.
    \item We use a dataset with binaries from 12 ISAs and show that byte-level $(1,2,3)$-gram TF-IDF features yield high accuracy (99\%) compared to the existing byte-histogram and  signature-based features (91\%).
    \item We show that character-level $(1,2,3)$-gram TF-IDF features extracted from encoded binaries yield high accuracy with $16 \times$ fewer features compared to the number of byte-level $(1,2,3)$-gram TF-IDF features. 
\end{itemize}

The remainder of this paper is organized as follows. Section~\ref{sec:prelim} provides the preliminaries on binary files and $N$-gram TF-IDF feature model from NLP. Section~\ref{sec:related} presents  related work. 
Section~\ref{sec:methods} presents the proposed NLP and encoding-based object code feature selection methods.
Sections~\ref{sec:experiments} and~\ref{sec:results} details the experiments and presents the experimental results. Section~\ref{sec:conclusion} provides concluding remarks.


\section{Preliminaries}\label{sec:prelim}

In this section we present  preliminaries on binary files and some  characteristics of  binaries that can be leveraged to identify the ISA. 

\subsection{Structures of Binary Files and Instruction Sets}\label{subsec:binary files}

We first detail the relevant components of the binary files studied in this paper. Then we introduce the concept of an ISA and two components of a binary instruction: opcode and operand.
\newline

\noindent
\textbf{A binary file} consists of instructions and resources (e.g., data values, memory addresses, file meta-data such as file size) that are stored as ``0"s and ``1"s (binaries).  Typically binaries are structured as 8-bit (1-byte) terms (e.g., 11010111 01000011). An 8-bit binary term is often represented as a two-digit Hexadecimal (Hex) number for ease of analysis (e.g., 11010111 01000011 in binary = d7 43 in hex). In this paper we focus on binary files corresponding to executable files, compiled programs, and operating system files that can be processed and executed by a computer to carry out tasks (e.g., computer programs). 

File header (e.g., ELF header, PE header) and object code are two important components of binary files required for their analysis. File header provides file meta-data such as file size, instruction length (e.g., 16-bit, 32-bit, 64-bit), details of any object code sections, and instruction set architecture of the processor that runs the binaries. The object code section contains binaries related to set of instructions. In this paper we assume file headers of the binary files are missing and only (partial) object code of binary files are available for analysis.
\newline

\begin{figure*}[!htb]
\includegraphics[width=11.5 cm]{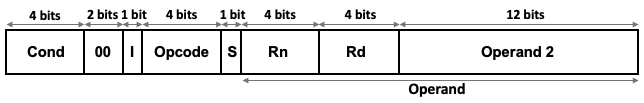}
\caption{{\color{black} 32-bit data processing instruction format of ARM architecture \cite{ARM}. Cond: Condition field, I: Immediate operand, Opcode: Operation code, S: Set condition codes, Rn: 1st operand register, Rd: Destination register. Typically, the opcode which defines the arithmetic and data operations (e.g., Add, Store)  to be carried out by the processor occupies fewer bits in an instruction compared to the operand which specifies the data values and memory/register addresses.}}\label{fig:ARM}
\end{figure*}

\begin{figure*}[!htb]
\includegraphics[width=11.5 cm]{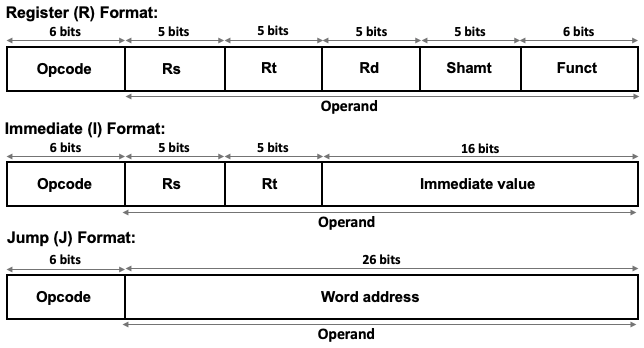}
\caption{{\color{black}Three different 32-bit instruction formats of MIPS architecture \cite{mips}. Rs: Source register, Rt: Source/destination register, Rd: Destination register, Shamt: Shift values, Funct: Instructions to harware, Immediate value: Stores the constant values used in the immediate instructions (e.g., ADDI: add immediate).
R format is used for the most arithmetic and logic instructions (e.g., ADD, XOR). I format is used for the data transfer, immediate and conditional branch instructions (e.g., MOVE, ADDI, BEQ: branch on equal). J format is used for unconditional jump instructions (e.g., JMP).  }  }\label{fig:MIPS}
\end{figure*}

\begin{figure*}[!htb]
\includegraphics[width=17.1cm]{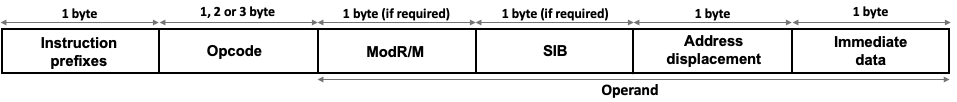}
\caption{{\color{black}Variable length instruction format of X86\_64 architecture \cite{x86-64-ins}. ModR/M: used for memory addresses or opcode extensions, SIB: Scaled index byte. The length of an X86\_64 instruction depends on the size of opcode or the usage of ModR/M and SIB. }}\label{fig:X86-64}
\end{figure*}

\noindent
\textbf{Instruction Set Architecture (ISA)} specifies rules for interpreting instructions in object code binaries to the processor. Examples of popular ISAs include ARM, MIPS, and x86\_64. The instruction length of an ISA indicates the number of consecutive bytes that defines one instruction for the processor. Architectures such as ARM and MIPS support fixed length (mostly 32-bit) instructions (Fig.~\ref{fig:ARM} and Fig.~\ref{fig:MIPS}) whereas x86\_64 supports variable length instructions (Fig.~\ref{fig:X86-64}). 
\newline

\noindent\textbf{Opcode and operand} are two main parts of a binary instruction. The opcode specifies the data transfer, arithmetic and logic, control, and floating point operations that need to be carried out by the processor. Length of an opcode varies both between and within different architecture types (e.g., 4-bit ARM opcodes in Fig.~\ref{fig:ARM} vs. 6-bit MIPS opcodes in Fig.~\ref{fig:MIPS} vs. 1,2 or 3-byte x86\_64 opcodes in Fig.~\ref{fig:X86-64}).  The operand  specifies the data values and memory/register addresses that are used in the operations specified by the opcode. 
The length of an operand depends on both architecture type and the type of data that it stores (e.g., 4-bit ARM destination register (Rd) address in Fig.~\ref{fig:ARM} vs. 5-bit MIPS R format Rd address in Fig.~\ref{fig:MIPS} vs. 26-bit MIPS J format memory address in Fig.~\ref{fig:MIPS}). Architectures have multiple instruction formats to support operations that use different type and number of operands (e.g., R, I and J instruction formats in Fig.~\ref{fig:MIPS}).

\subsection{Endianness of an Instruction Set  Architecture}\label{subsec:endianness}

The endianness defines the \textit{byte order} of a multi-byte data variable representation. There are two types of endianness: (i)~little-endian and (ii)~big-endian. Under little-endian representation of a 2-byte number, the most-significant byte occupies the lowest memory address. In big-endian representation, the least-significant byte occupies the lowest memory address. For example, a data variable with value 1 is stored as  0x0100 and  0x0001 under little-endian and big-endian 2-byte representations, respectively. Examples of common little-endian ISAs include ARM, AVR, and x86\_64. ISAs such as MIPS, PowerPC, and SPARC use big-endian representation.



\subsection{$N$-gram Term Frequency-Inverse Document Frequency}\label{subsec:TF-IDF}

 {\color{black} $N$-gram, Term Frequency (TF), and Inverse Document Frequency (IDF) are widely used feature methods in Natural Language Processing (NLP) tasks such as text mining and auto-completion of sentences \cite{ramos2003using, aizawa2003information, trstenjak2014knn, bafna2016document}. We provide the definitions and discuss some of the characteristic properties of these NLP-based feature methods below.}
\newline

\noindent
\begin{defn}[\textbf{$N$-gram}]
 An $N$-gram is a contiguous sequence of $N$ terms (e.g., characters, words) in the content of a document (e.g., text message, news article).  Typically, $N$-grams are extracted by moving a window of length $N$ forward, one\footnote{NLP applications that involves learning subject-verb relationships and word semantics in languages may require moving the window forward by multiple terms.} term at a time, along the content of a document. 
 \newline
\end{defn}

\begin{defn}[\textbf{Term Frequency} \cite{ramos2003using}]
 The term frequency (TF) is the number of times each term appears in a document. Typically, TF values are divided by the total number of terms (normalized) in the document to mitigate the effect of the document length\footnote{Documents can have different lengths with regard to the number of terms recorded in them. In such scenario, a term $\tau$ may appear more frequently in longer documents compared to shorter documents.  Hence unnormalized  TF values of $\tau$ in longer documents will be higher than the values corresponding to shorter ones faulty indicating $\tau$ is more dominant pattern inherent to longer documents.}. TF of a term $\tau$ is
\begin{equation}\label{eqn:TF}
    \mbox{TF($\tau$)} = \frac{\mbox{\# of times $\tau$ appeared in the document}}{\mbox{\# of terms in the document}}. 
\end{equation}
\end{defn}

\noindent
\begin{defn}[\textbf{Inverse Document Frequency} \cite{ramos2003using}]
The inverse document frequency (IDF) measures the informativeness of terms in a collection of documents (corpus).
It assigns lower values to terms that commonly appear among the documents in the corpus as they do not contribute in distinguishing the contents of documents. Conversely, higher values are assigned to the less frequent terms in the corpus as they may constitute patterns inherent to the content of the documents. 
IDF of a term $\tau$ is
\begin{equation}\label{eqn:IDF}
    \mbox{IDF($\tau$)} = \log \left(\frac{\mbox{\# of documents in the corpus} + 1}{\mbox{\# of documents with $\tau$} + 1} \right) + 1.
\end{equation}
\newline
\end{defn}

\noindent
\begin{defn}[\textbf{TF-IDF} \cite{ramos2003using}]
 The TF-IDF is a statistic that measures the importance of a term to a document in a corpus. TF-IDF value of a term $\tau$ is 
defined as:
\begin{equation}\label{eqn:TF-IDF}
    \mbox{TF-IDF($\tau$)} = \mbox{TF($\tau$)} \times \mbox{IDF($\tau$)}
\end{equation}
\newline
\end{defn}
In the context of NLP tasks, the $N$-gram TF-IDF presents a way to associate meaningful numerical values to words (i.e., scoring of words) that can be provided as the inputs to the ML models. A number assigned to a $N$-gram word in a document under $N$-gram TF-IDF method is increased proportionally by the number of times the corresponding $N$-gram word appears in the particular document, but is decreased by the number of documents that contain the word. Therefore,  words that are common in each document, such as ``a", ``the", ``and", ``this", ``that", and ``if" will have lower TF-IDF values even though they may appear frequently, as they do not help in identifying the content of a document in particular. However, if a specific $N$-gram word appears frequently in a document (or a smaller subset of documents), while not appearing frequently in others, it probably means that this particular $N$-gram word is very relevant to the content of the document (or the subset of documents). {\color{black} For example, if a $1$-gram word ``Computer" appears frequently in $20$ web articles out of $1000$, then it is likely that these $20$ articles are related to computers.} 

\subsection{Encoded Binaries}\label{subsec:encoded binaries}

A binary-to-text encoding converts  binary data to a sequence of printable characters. Such encodings are necessary for transmission of data when the communication channels do not allow binary data (such as email or Network News Transfer Protocol-NNTP). Rows~III-VI in Table~\ref{table:Encoding} below present examples of the binary code in Row~I encoded into four popular binary-to-text encoding methods; Base16, Base32, Base64, and Base 85, respectively. 

Under Base16, every byte of binary data is encoded to 2 characters. Base32 encodes 5-bytes of binary data into 8 characters. Base64 encodes 3-bytes of binary data into 4 characters. Lastly, Base85 encodes 4-bytes of data into 5 characters. The suffix of the encoding method name provides the number of distinct characters included in the alphabet of the encoding method. For example Base16 alphabet consists of 16 characters; digits $0-9$, and letters $A-F$. Additionally, Base32 and Base64 encoding use a padding character, ``$=$".

\begin{table}[!h]
\resizebox{8.25cm}{!}{
\begin{tabular}{l|l|l|l|}
\cline{2-4}
    & \multicolumn{2}{c|}{\textbf{File format}}    & \multicolumn{1}{c|}{\textbf{Example}}                                                                              \\ \cline{2-4} 
I   & \multicolumn{2}{l|}{\textbf{Binary}}                  & \begin{tabular}[c]{@{}l@{}}11010111 01000011 11010100 01000100 \\ 11010110 01000100 11011000 01000101\end{tabular} \\ \cline{2-4} 
II  & \multicolumn{2}{l|}{\textbf{Hex}}                     & d7 43 d4 44 d6 44 d8 45                                                                                            \\ \cline{2-4} 
III & \multicolumn{2}{l|}{\textbf{Base16 encoded}} & D743D444D644D845                                                                                                   \\ \cline{2-4} 
IV  & \multicolumn{2}{l|}{\textbf{Base32 encoded}} & 25B5IRGWITMEK===                                                                                                   \\ \cline{2-4} 
V   & \multicolumn{2}{l|}{\textbf{Base64 encoded}} & 10PURNZE2EU=                                                                                                       \\ \cline{2-4} 
VI  & \multicolumn{2}{l|}{\textbf{Base85 encoded}} & f0e\%UejS.Z                                                                                                         \\ \cline{2-4} 
\end{tabular}
}
\caption{An example illustrating different binary files formats. Binary data in row~I has been grouped into 8-bit values to represent 1-byte of data following the normal convention. Row~II shows the hexadecimal (Hex) representation of the binaries in row~I. Row~III, Row~IV, Row~V, and Row~VI present the Base16, Base32, Base64, and Base85 encoding of the binaries given in Row~I, respectively.}\label{table:Encoding}
\end{table}



\section{Related Work}\label{sec:related}


In this section we first present an overview of binary code feature extraction techniques proposed in the literature for different applications. Then we present existing binary code features used for ML-based ISA identification and provide a brief review of cyber security applications that use the N-gram TF-IDF feature model.

Feature extraction from binaries has been widely studied in the areas of ML-based malware detection \cite{kolter2006learning, saxe2015deep, schultz2000data} and file type identification \cite{mcdaniel2003content, karresand2006file, ahmed2010content}. The authors of \cite{kolter2006learning} presented a feature extraction method for malware detection that uses information gain to select top 500, 4-gram byte patterns found in the training set binaries. The work in \cite{saxe2015deep} proposed malware detection using a set of features composed with byte entropy histograms, string 2D histograms, and vectors corresponding to the hash values of binary file's metadata and import address tables. The authors of \cite{schultz2000data} combined the Boolean features related to the usage of dynamic-link library (DLL) files, function calls, GNU strings, and byte sequences to detect malware. The work in \cite{mcdaniel2003content} introduced byte frequency distributions (BFDs) of object code and file header/trailer, and byte frequency cross-correlation distribution for file type identification. The authors of \cite{karresand2006file} suggested file type identification using a combination of the BFD and the frequency distribution corresponding to the rate of change of the byte content. The authors of \cite{ahmed2010content} used cosine similarity of BFD features to predict the file type.

ML frameworks presented in \cite{sickendick2013file} and \cite{ma2019svm} use information gain based top $N$ 4-gram byte features for target architecture identification of firmware binaries. However, these approaches assume that a single instruction is stored in four bytes (32-bit) and hence are not suitable when identifying architectures that use different byte sizes (e.g., 8-bit, 16-bit, 64-bit)  or variable length instructions (e.g., x86\_64) to store instructions. Moreover, information gain computations require  finding the number of different instruction set bit patterns recorded in each 4-gram byte pattern corresponding to each target architecture considered. Such a pattern searching procedure becomes computationally exhaustive when the number of target architectures is increased. 






The author of \cite{clemens2015automatic} proposed to combine the byte-histogram features (i.e., BFD) with a set of features that capture endianness of binaries for ISA identification. Under this method, all the features are extracted from the decoded program binaries (files in binary or hex format). First for each binary file, a normalized byte-histogram is generated by counting all individual byte values in the file content. This provides $256$ features. The endianness features are extracted by counting byte pairs which correspond to code sections –
which increment by one (0x0001 vs 0x0100) – as well as those sections that correspond to a decrement by one
(0xfffe vs 0xfeff). 
{\color{black}
The authors of \cite{beckman2020binary} adapted the byte-histogram along with endianness features introduced in \cite{clemens2015automatic} and proposed a simplified technique to determine the endianness of a binary file.  Under the simplified endianness features, if it was found that there were more 0x0001’s than 0x0100’s then the entry of the feature vector corresponding to the big endianness was assigned a value of 1 and the entry corresponding to the little endianness was assigned 0. If it is found that the abundance is of the form 0x0100, then the reverse assignments are made.

Recently, the works \cite{nicolao2018elisa} and \cite{kairajarvi2020isadetect} extended the byte-histogram and endianness features introduced in \cite{clemens2015automatic} by adding additional signature-based features extracted from the function epilogue and function prologue sections. Specifically, \cite{nicolao2018elisa} introduced 31 new signature-based features extracted from binary files of amd64, arm, armel, mips32, powerpc, powerpc64, s390x and x86 ISAs. The authors of \cite{kairajarvi2020isadetect} introduced two additional signature-based features to identify powerpc ISAs. However, there is no guarantee that these signature-based features are included in partial binaries. As observed in \cite{nicolao2018elisa}, extending signature-based features to identify additional ISAs will require significant effort and expert knowledge. 
Additionally, we note that the byte-histogram or BFD features are susceptible to frequent byte patterns commonly observed among binaries of different architectures (i.e., noisy byte patterns).

The N-gram TF-IDF feature model has been used in cyber security applications such as software vulnerability assessments \cite{behl2014bug, yuan2020detection} and cyber threat detection \cite{chen2008intrusion, tran2017nlp, zhang2019classification, ali2020malgra}. The authors of \cite{behl2014bug} used TF-IDF features extracted from bug reports to develop a tool for identifying software bugs. The work presented in \cite{yuan2020detection} used TF-IDF features of Android application package’s (Apk’s) manifest file to evaluate the security of Android applications. The authors of \cite{chen2008intrusion} extracted TF-IDF features from process logs to build an intrusion detection system for a computer network. The research presented in \cite{tran2017nlp} used TF-IDF features extracted from opcode sequences to classify ransomware families. The authors of \cite{zhang2019classification} and \cite{ali2020malgra} used TF-IDF features extracted from Application Programming Interface (API) call sequences for malware classification.
}




\begin{table*}[b]
\resizebox{11.8cm}{!}{
\begin{tabular}{|p{2.7cm}|p{2cm}|p{1.2cm}|p{1.2cm}|p{1.2cm}|p{0.3cm}|}
\hline
\multicolumn{1}{|c|}{\multirow{2}{*}{\textbf{\begin{tabular}[c]{@{}c@{}}Binary\\ file format\end{tabular}}}} & \multirow{2}{*}{\textbf{\begin{tabular}[c]{@{}c@{}}Granularity of \\  features\end{tabular}}} & \multicolumn{4}{c|}{\textbf{Number of features}}                                          \\ \cline{3-6} 
\multicolumn{1}{|c|}{}                                                                                       &                                                                                               & \textbf{1-gram} & \textbf{2-gram} & \textbf{3-gram} & \multicolumn{1}{c|}{\textbf{Total}} \\ \hline
\textbf{Base16 encoded}                                                                                      & character                                                                                     & 16              & 256             & 4096            & \textbf{4368}                       \\ \hline
\textbf{Base32 encoded}                                                                                      & character                                                                                     & 32              & 1024            & 5000            & \textbf{6056}                       \\ \hline
\textbf{Base64 encoded}                                                                                      & character                                                                                     & 64              & 4096            & 5000            & \textbf{9160}                       \\ \hline
\textbf{Base85 encoded}                                                                                      & character                                                                                     & 85              & 7225            & 5000            & \textbf{12310}                      \\ \hline
\textbf{Decoded (in Hex)}                                                                                    & byte                                                                                          & 256             & 65536           & 5000            & \textbf{70792}                      \\ \hline
\end{tabular}
}
\caption{Details of the (1,2,3)-gram TF-IDF features extracted from the different encoded binary file formats. }\label{table:TF-IDF}
\end{table*}
\section{Proposed NLP Approaches for Object Code Feature Extraction}\label{sec:methods}

In this section we propose two object code feature selection methods named byte-level (1,2,3)-gram TF-IDF features (Section~\ref{subsec:byte-level}) and encoded character-level (1,2,3)-gram TF-IDF features (Section~\ref{subsec:char-level}) for ML-based ISA identification. We first present the key observations that motivate the $N$-gram TF-IDF structure of the two object code feature methods.

The accuracy of instruction set architecture identification largely depends on the ability of the set of object code features to capture  bit patterns that help distinguish between different architectures. In Section~\ref{subsec:binary files}, we observed that the opcodes and operands are two of the most important information embedded in binary instructions and they can have different lengths both within the instructions of same ISA and across the instructions of different ISAs. Hence, identifying a sufficient number of bit patterns that can enable high accuracy ISA identification across increased number of architectures is a non-trivial task that requires domain knowledge about the ISAs.


In what follows, we describe a method to characterize binary code features that does not require domain knowledge about the ISAs. 
Our approach involves the following steps: i) identify the smallest unit that has meaning (e.g., in NLP, this unit is a word) within the binaries; ii) identify the fixed-length patterns that need to be extracted from the binaries; iii) form a frequency vector of all possible lengths $1, 2, \dots$ that can be used as the set of features for the binary files. 
%
We note that the frequency vector must satisfy the following properties:
\begin{enumerate}
    \item Length of object code binaries should not influence\footnote{
Consider any two binaries $X$ and $Y$ of same architecture type $A$ with $X$ having smaller length and $Y$ having significantly larger object code length. In such scenario, $X$ will be seen as much less type $A$ compared to $Y$ as the frequency of the architecture specific bit patterns in $X$ will be always significantly less than the frequency of the same bit patterns in $Y$.} frequency values.
    \item Values corresponding to frequent patterns that also commonly appear among the binary files (i.e., noisy patterns) should be attenuated as they do not help in distinguishing between the ISAs of the binaries.
    \item Values corresponding to frequent patterns appearing only in a small subset of binary files should be boosted since such patterns will have higher probability of being ISA-specific patterns that can aid architecture identification.
\end{enumerate}


We adapt $N$-gram TF-IDF feature model for extracting object code features. We propose two approaches to determine what will constitute a meaningful \emph{term} in binaries and selecting an appropriate $N$ for capturing architecture prominent patterns.


\subsection{Byte-level $N$-gram TF-IDF Features}\label{subsec:byte-level}

As noted in Section~\ref{subsec:binary files}  an instruction recorded in a binary file is composed of a collection of consecutive bytes. Processors read and process each code section of a binary file byte-by-byte to execute instructions. Therefore, we first choose a byte as a term when adapting $N$-gram TF-IDF for extracting features from object code binaries. 

Opcodes define architecture specific operations whereas operands define data and addresses that usually takes random values. Thus, the opcodes consist of more structured patterns that can characterize ISAs of binaries. For example in Section~\ref{subsec:binary files} we observed that ARM and MIPS architectures typically have opcodes of length of 4-bits and 6-bits (Fig.~\ref{fig:ARM} and Fig.~\ref{fig:MIPS}) respectively while opcodes of  X86\_64 can be either 1-byte, 2-byte or 3-bytes. This implies we can expect higher TF-IDF values for 1-gram byte patterns that include 4-bit and 6-bit opcode patterns in respective ARM and MIPS feature vectors. Similarly feature vectors of X86\_64 binaries will have higher TF-IDF values for 1-gram, 2-gram or 3-gram  byte patterns corresponding to 1-byte, 2-byte or 3-byte opcodes, respectively. Therefore, we extract $1$-gram, $2$-gram and $3$-gram (i.e., $N = 1, 2,$ and $3$) byte patterns from binaries and use a vector corresponding to their TF-IDF values as the binary object code features.

Further, 2-gram byte-level TF-IDF values allow capturing consecutive byte patterns in operands that are induced by the endianness property of an architecture (Section~\ref{subsec:endianness}). {\color{black}For example, the data value $1$ can be considered as a commonly used operand across the binaries as many object codes may include instructions related to increasing  \emph{for} and \emph{while} loops variables by 1.} Then TF-IDF feature vectors of big endian MIPS architecture binaries can expect to have higher values associated with the 0x0001 2-gram byte pattern compared to the values of 0x0100.  On the other hand features of little endian ARM and X86\_64 may have higher values for 0x0100 and relatively lower values for 0x0001.

Our proposed byte-level object code features include all $(1,2)$-gram byte patterns and top $5000$ ranked 3-gram byte-patterns. The rank ordering of the observed 3-gram byte-patterns are done using the frequency of those patterns across all binary files in the training set binaries. We only include top $5000$ ranked 3-gram byte-patterns since capturing all such patterns will require a large number of features ($256^3 >> 5000$) that can drastically increase the computation time and resources such as memory and processing power required for ML-based ISA identification. {\color{black}Therefore, byte-level (1,2,3)-gram TF-IDF features do not depend on a limited number of pre-selected byte patterns based on the domain knowledge and heuristics that may be completely absent in some (partial) binary data. Rather, our approach provides a more general set of expert agnostic features to identify byte patterns induced by opcodes and endianness of ISAs. 
} There are $2^8 = 256$ possible 1-gram byte (8-bit) patterns. Hence, the total number of required feature values to represent each binary file under this method can be as large as $256 + 256^2 + 5000 = 70792$.




\subsection{Character-level $N$-gram TF-IDF Features from Encoded Binaries}\label{subsec:char-level}

As discussed in Section~\ref{subsec:byte-level}, bit-patterns of opcodes play a more vital role in identifying ISAs since they contain operations inherent to the target processor architecture. In addition, we observe that opcode information embedded in the instructions are typically less than 1-byte (e.g., 4-bit opcodes in ARM, 6-bit opcodes in MIPS, AVR, and PowerPC) with the exception of 1,2, or 3-byte opcodes in X86\_64. {\color{black}We also observe that there are other  bit patterns of  different lengths  recorded in instructions (outside the bit patterns related to opcodes and operands) that may be used to distinguish between architecture types. Examples include 4-bit condition field in ARM instructions which is (Fig.~\ref{fig:ARM}) mostly set to $1110$ for indicating ``always execute" and 1-byte instruction prefixes in X86\_64 instructions (e.g., 0xf0 - repeat/lock prefix; 0xf2 and 0xf3 - string manipulation prefixes). Hence, accurately capturing such fine-grained bit patterns specific to architectures requires choosing terms with less than 8-bits (ideally 4, 5 or 6-bit terms). }

As noted in Section~\ref{subsec:encoded binaries}, encoding methods naturally provide a way to group binary data into different length bits. Therefore, we propose to extract character-level (1,2,3)-gram TF-IDF features from encoded binaries to enable capturing architecture specific fine grained bit patterns. Table~\ref{table:TF-IDF} provides the number of features we used under each encoding method and their composition. Using this method allows us to reduce the required number of features by approximately $16 \times$ compared to byte-level features discussed in Section~\ref{subsec:byte-level}.






\begin{table*}[t]
\resizebox{11.8cm}{!}{
    \centering\noindent
    \begin{tabular}{|p{2.6cm}|p{1.4cm}|p{1.3cm}|p{1.4cm}|p{1.5cm}|p{1cm}|}
\hline
\multicolumn{1}{|c|}{\multirow{2}{*}{\textbf{\begin{tabular}[c]{@{}c@{}}Binary\\ file format\end{tabular}}}} & \multicolumn{2}{c|}{\textbf{Granularity of features}} & \multicolumn{3}{c|}{\textbf{Number of features}}          \\ \cline{2-6} 
\multicolumn{1}{|c|}{}                                                                                       & \textbf{Histogram}        & \textbf{Endianness}       & \textbf{Histogram} & \textbf{Endianness} & \textbf{Total} \\ \hline
\textbf{Base16 encoded}                                                                                      & character                 & 2-byte                    & 16                 & 4                   & \textbf{20}    \\ \hline
\textbf{Base32 encoded}                                                                                      & character                 & 2-byte                    & 32                 & 4                   & \textbf{36}    \\ \hline
\textbf{Base64 encoded}                                                                                      & character                 & 2-byte                    & 64                 & 4                   & \textbf{68}    \\ \hline
\textbf{Base85 encoded}                                                                                      & character                 & 2-byte                    & 85                 & 4                   & \textbf{89}    \\ \hline
\textbf{Decoded (in Hex)}                                                                                    & byte                      & 2-byte                    & 256                & 4                   & \textbf{260}   \\ \hline
\end{tabular}
}
\captionof{table}{Details of histogram + endianness features extracted from different binary file formats.}\label{table:histogram}
\end{table*}

\section{Experiment Setup}\label{sec:experiments}

This section presents the details of the experiments used to compare the performance of ML algorithms for ISA identification under two types of feature selection methods: $1)$~Histogram along with Endianness (Hist. + Endian) features and $2)$ (1,2,3)-gram TF-IDF features. First, we detail the characteristics of the datasets used in our experiments. {\color{black}Then we present the properties of the different types of features extracted from the datastes. }  All the experiments are implemented using Python 3.8.5 on a workstation with Intel(R) Xeon(R) W-2145 CPU @ 3.70GHz processor and $128$~GB memory. 




\subsection{Datasets of Object Code Binaries Used}
Our primary dataset consists of $202,066$ distinct Base64 encoded binaries downloaded from Praetorian's ``Machine Learning Binaries" challenge web page \cite{praetorian_2021}. Each encoded binary string in this dataset consists of 88 characters (66 bytes) on average and belongs to one of the following twelve architecture types: avr, alphaev56, arm, m68k, mips, mipsel, powerpc, s390, sh4, sparc, x86\_64, and xtensa. We divided the primary dataset into 50 non-overlapping groups, each of which was further partitioned into a training set with $2856$ encoded binaries ($238$ per architecture) and a testing set with $960$ encoded binaries ($80$ per architecture). 

{\color{black}We applied a series of decoding and encoding operations to the Base64 encoded binaries of each dataset to create datasets for four other data formats: binary, Base16, Base32, and Base85. The 8-bit (byte) values in binary formatted datasets were further converted to their corresponding 2-character Hex values for  ease of  analysis.}



\subsection{Baseline Object Code Features}
{\color{black}
We use the byte-histogram and endianness features introduced in \cite{clemens2015automatic} for ISA identification as our baseline for byte-level (1,2,3)-gram TF-IDF features. The byte-histogram features of each binary are extracted by counting the number of times each distinct Hex value appears in the decoded binary. The byte-histogram of each binary is then normalized by the total number of Hex values recorded in the corresponding binary.  Note that the byte-histogram features are equivalent to the 1-gram TF features and they form the first $2^8 = 256$ entries in the feature vector. Then four more domain knowledge/heuristic-based features are added to the feature vector for capturing the endianness. These additional features are extracted by counting the number of times each 2-byte Hex values, 0x0001, 0x0100, 0xfffe, and 0xfeff appear in each binary. These counts are also normalized by the total number of Hex values in the binary.

In order to evaluate the effectiveness of our character-level (1,2,3)-gram TF-IDF features extracted from the encoded binaries, we use character-histogram and endianness features. The character-histogram features of each binary are extracted by counting the number of times each distinct character appears in the encoded binary. The character-histogram of each encoded binary is then normalized by the total number of characters recorded in the corresponding binary. Since the domain knowledge/ heuristic-based four endianness features introduced in \cite{clemens2015automatic} are based on the 2-byte Hex values, we extract the four endianness features from the decoded binaries in Hex format following the same steps as in byte-histogram and endianness features.  Table~\ref{table:histogram}  summarizes the details about the \textit{histogram + endianness} features.
}

\section{Results and Discussions}\label{sec:results}

In this section, we present the experimental results and related discussions.  We use the following abbreviations to denote the different ML algorithms used in the experiments. SVM: Support Vector Machine, LR: Logistic Regression, DT: Decision Tree, RF: Random Forest, GNB: Gaussian Naive Bayes, MNB: Multinomial Naive Bayes, CNB: Complement Naive Bayes, KNN: K-Nearest Neighbor, and PTN: Perceptron.   We will use \textit{histogram + endianess} to refer to the baseline features. We refer to \cite{harrington2012machine} for the detailed descriptions about the aforementioned ML algorithms.

\begin{figure*}[h]
    \centering\noindent
    {\includegraphics[width=0.55\linewidth,scale=1]{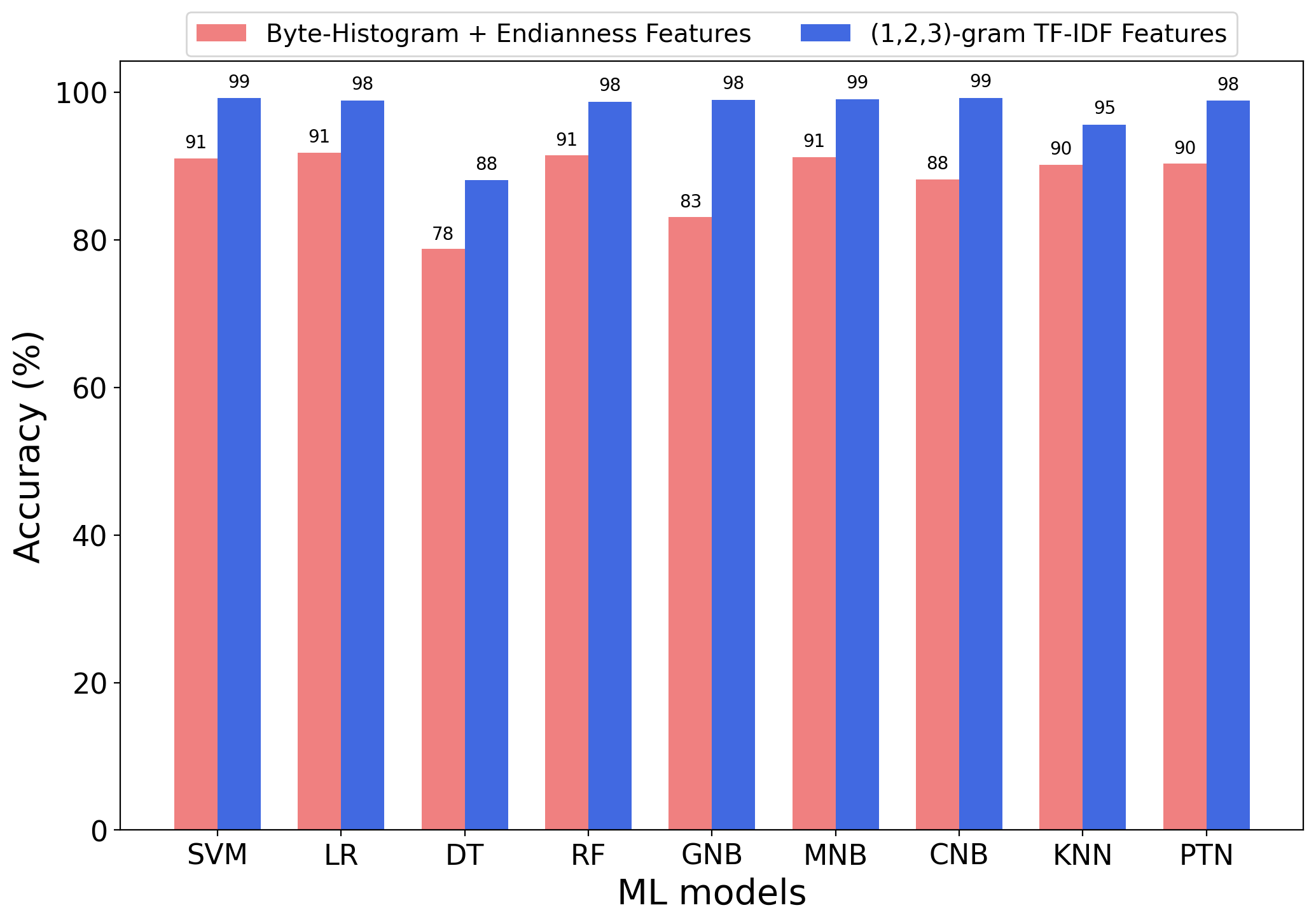}%
}
    \captionof{figure}{Accuracy of instruction set architecture identification under different machine learning algorithms corresponding to byte-histogram along with endianness features and byte-level (1,2,3)-gram TF-IDF features extracted from decoded binaries. Each accuracy value is computed across 50 independent datasets. The byte-level (1,2,3)-gram TF-IDF features consistently results in a higher accuracy compared to the byte-histogram + endianness features.  }
    \label{fig:decoded}
\end{figure*}

\begin{figure*}[!h]
\floatsetup{capposition = below, floatrowsep =qquad,}
\centering
\ffigbox{%
\begin{subfloatrow}
\centering
\ffigbox[0.45\textwidth]{\caption{Base16 encoded binaries}}{%
\includegraphics[width=82mm,scale=0.7]{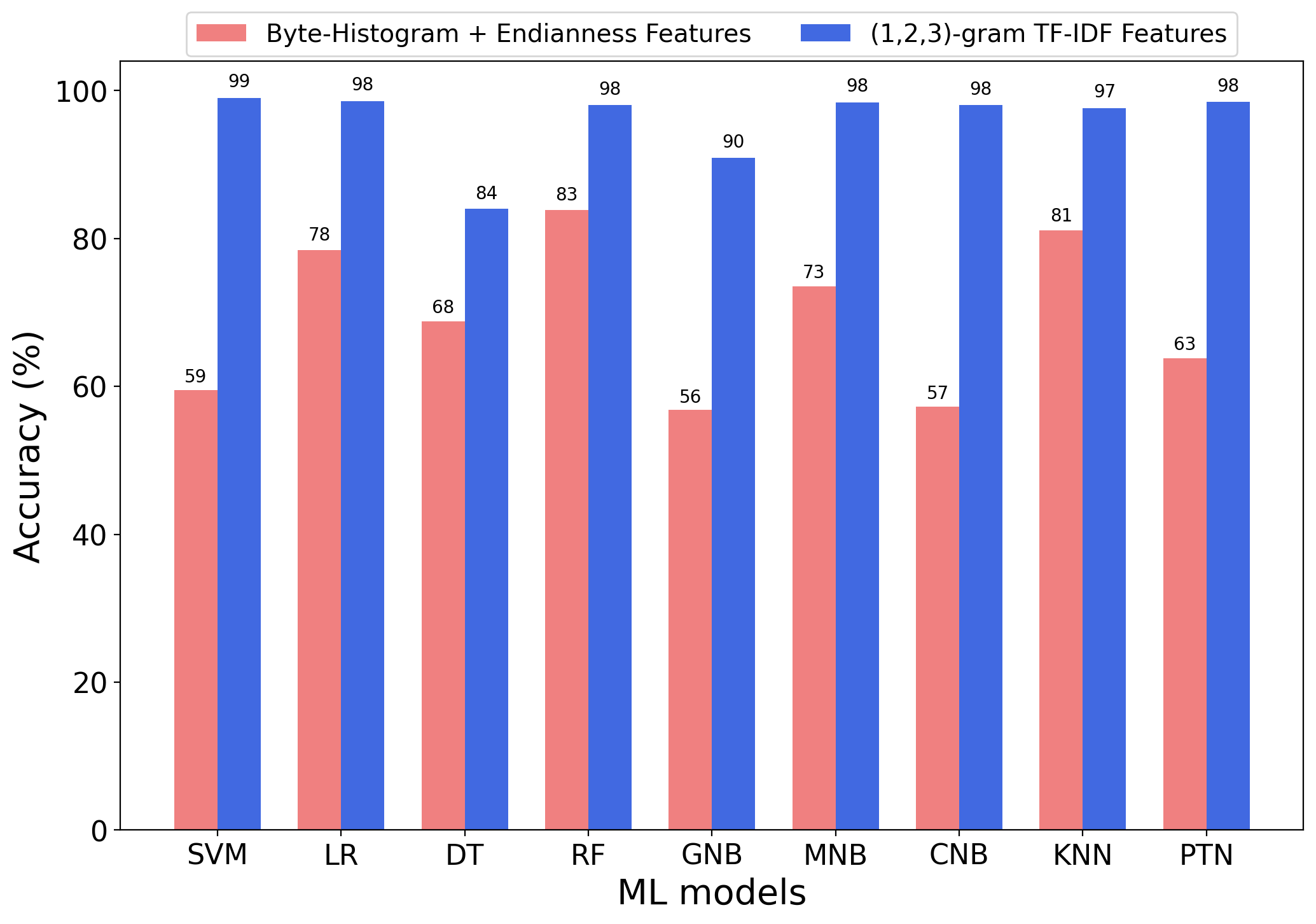}
}
\ffigbox[0.45\textwidth]{\caption{Base32 encoded binaries}}{%
\includegraphics[width=82mm,scale=0.7]{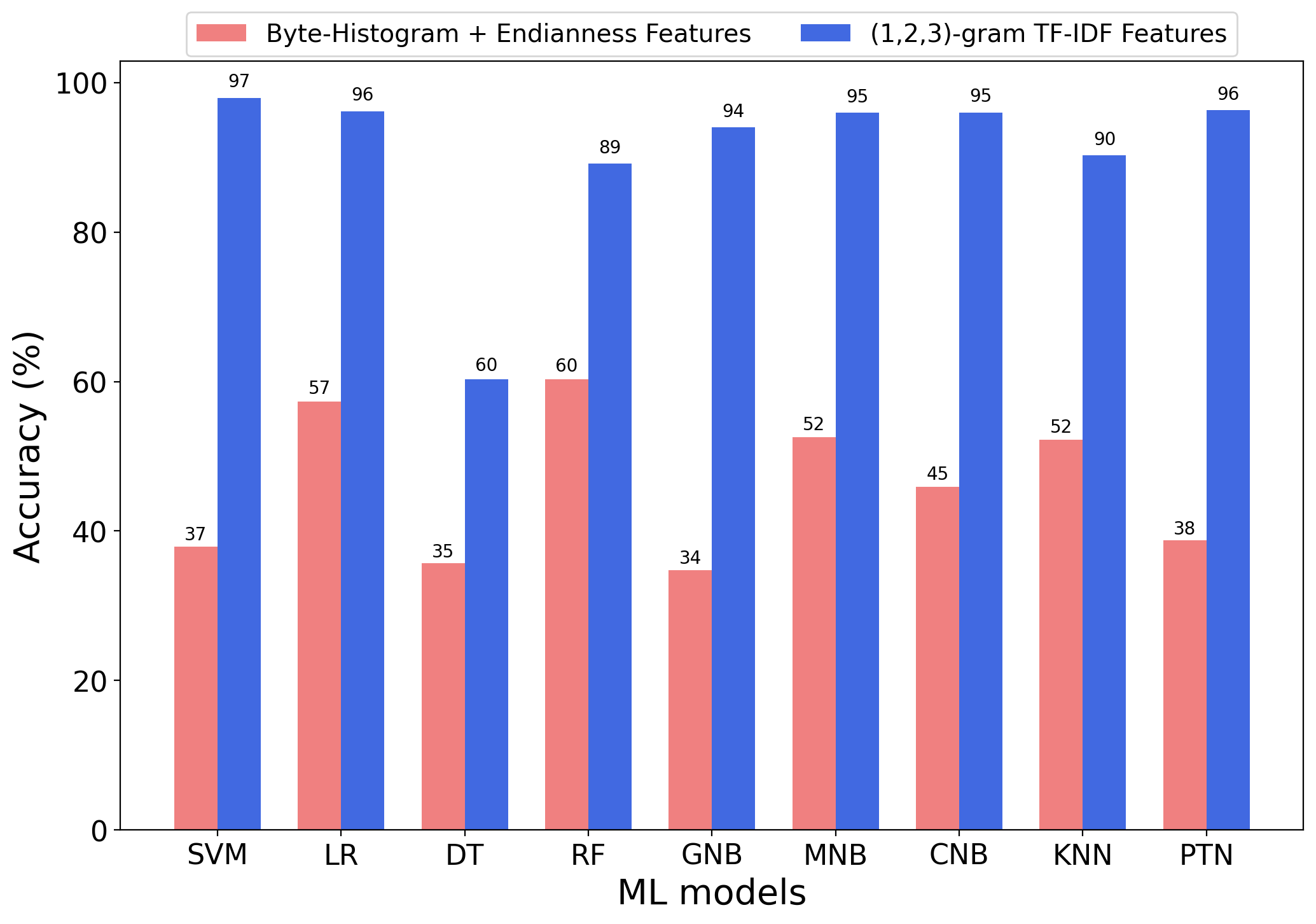}
}
\end{subfloatrow}
\vskip4ex
\begin{subfloatrow}
\ffigbox[0.45\textwidth]{\caption{Base64 encoded binaries}}{%
\includegraphics[width=82mm,scale=0.7]{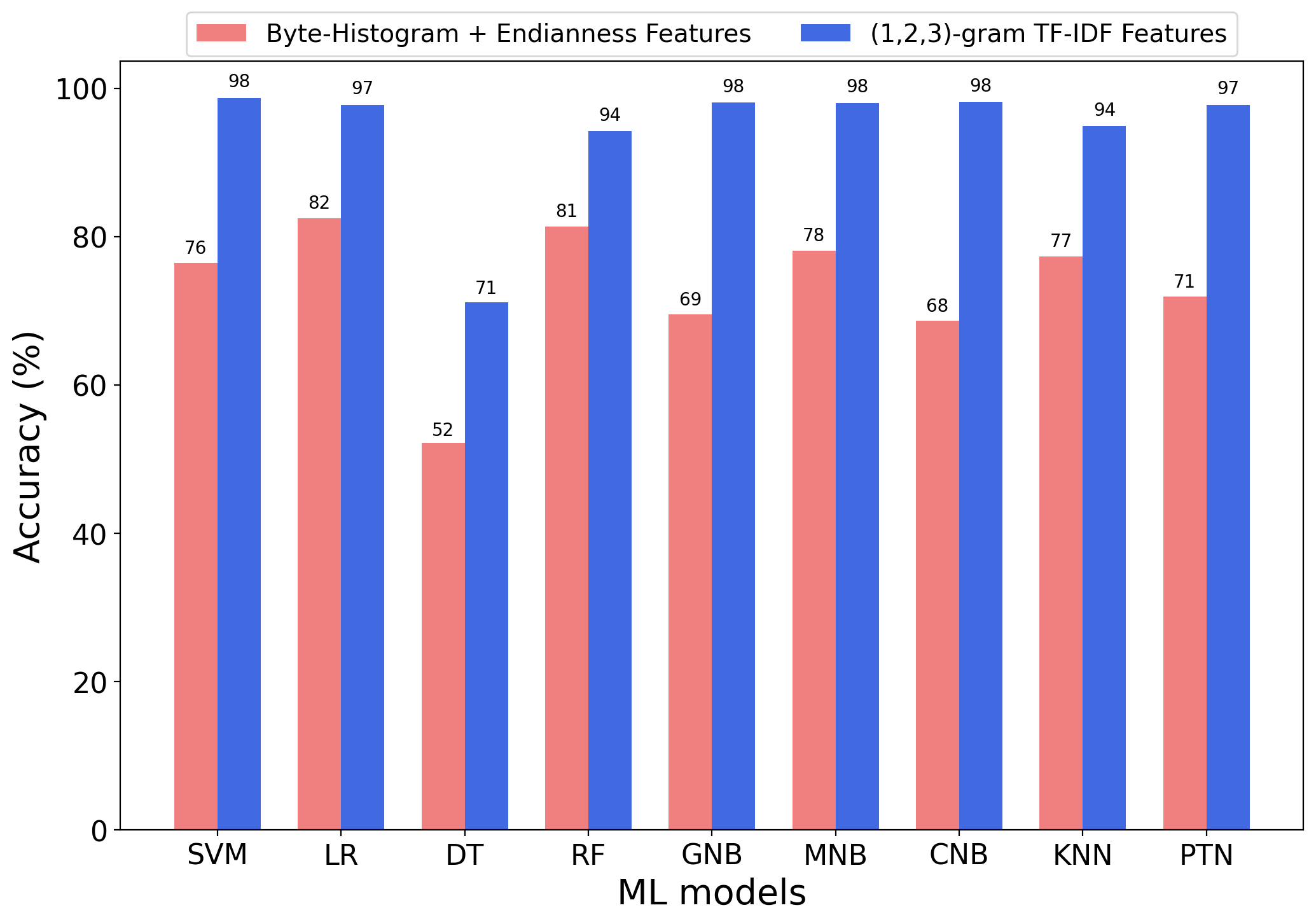}%
}
\ffigbox[0.45\textwidth]{\caption{Base85 encoded binaries}}{%
\includegraphics[width=82mm,scale=0.7]{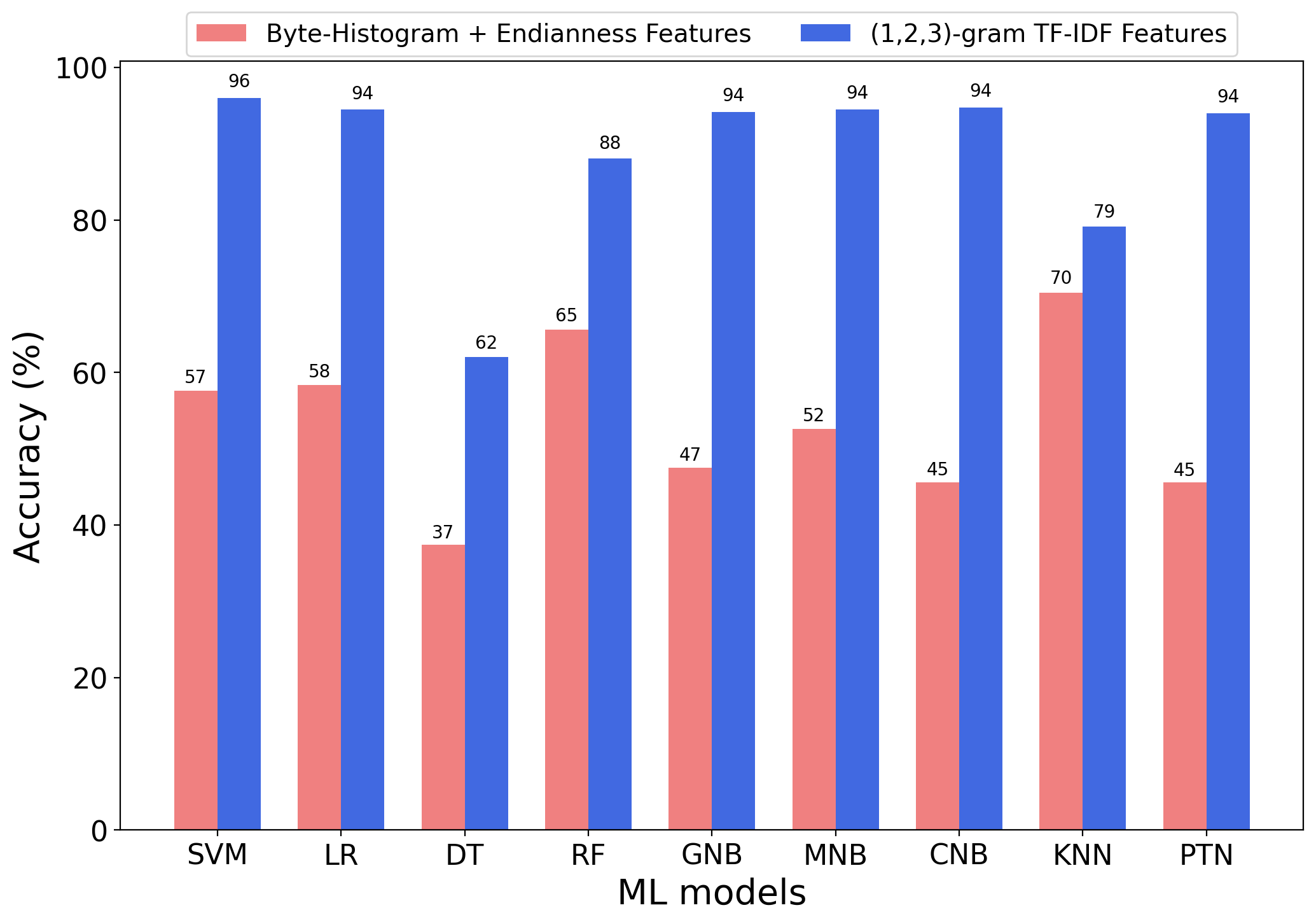}%
}
\end{subfloatrow}
}{
\caption{Accuracy of 12 architecture program binaries classification of different machine learning algorithms corresponding to character-histogram along with endianness features and character-level (1,2,3)-gram TF-IDF features under different encoded binary file formats, Base16, Base32, Base64, and Base85. Each accuracy value is computed across 50 independent datasets. The character-level (1,2,3)-gram TF-IDF features consistently results in a higher accuracy compared to the histogram + endianness features. }
\label{fig:encoded}}
\end{figure*}
\begin{figure*}[b]
\floatsetup{capposition = below, floatrowsep =qquad,}
\centering
\ffigbox{%
\begin{subfloatrow}
\centering
\ffigbox[0.45\textwidth]{\caption{Byte-histogram + Endianness features from decoded binaries}}{%
\includegraphics[width=77mm,scale=0.8]{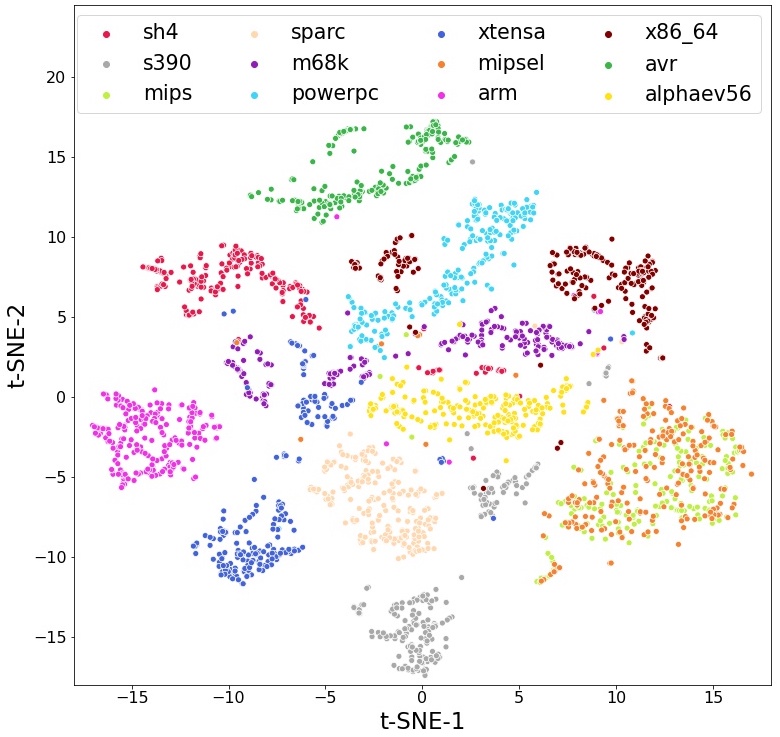}
}
\ffigbox[0.45\textwidth]{\caption{Byte-level (1,2,3)-gram TF-IDF features from decoded binaries}}{%
\includegraphics[width=77mm,scale=0.8]{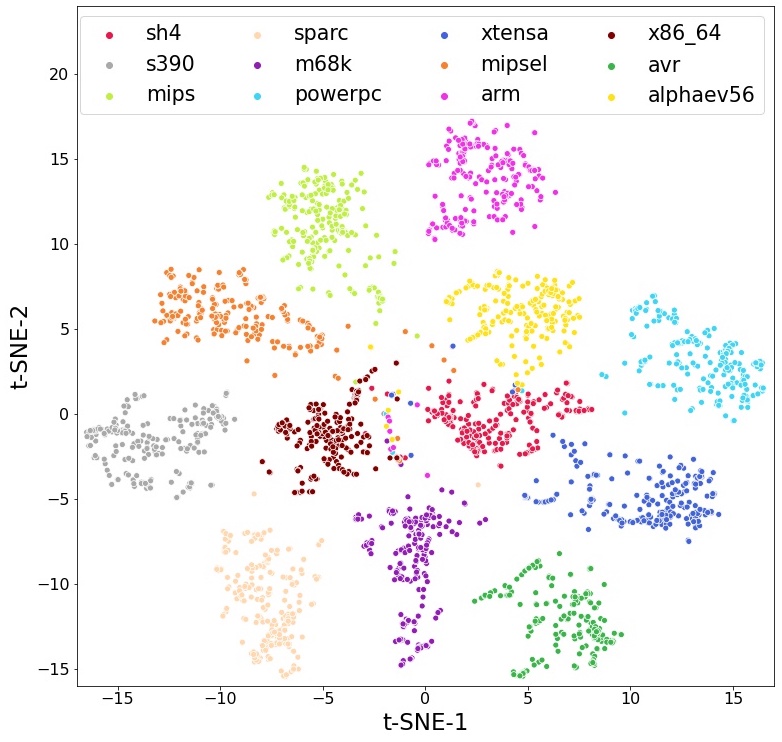}
}
\end{subfloatrow}
}{
\caption{2-D t-SNE plots corresponding to byte-histogram + endianness features (Fig.~\ref{fig:tsne-decoded}-(a)) and byte-level N-gram TF-IDF features (Fig.~\ref{fig:tsne-decoded}-(b)) extracted from randomly chosen training datasets consist of decoded binaries. The byte-level (1,2,3)-gram TF-IDF features provide better separability compared to the byte-histogram + endianness features.}
\label{fig:tsne-decoded}}
\end{figure*}

\subsection{Accuracy of the ML models}

Fig.~\ref{fig:decoded} compares the accuracy of instruction set architecture identification under different ML algorithms corresponding to {byte-histogram + endianness} features and byte-level (1,2,3)-gram TF-IDF features extracted from the decoded binaries. 
Accuracy values that we report in our experiments are averaged over 50 independent datasets that contain binaries from 12 different architectures.
Our results show that using byte-level (1,2,3)-gram TF-IDF features increase the accuracy values  by $9\%-10\%$ on average across the ML models considered. {\color{black}Properties of byte-level (1,2,3)-gram TF-IDF features such as ability to suppress the effects of noisy bytes, providing more generalized set of features to capture the architecture related characteristics such as endianness, and providing increased number of features that can capture byte patterns specific to architectures enables consistently achieving higher levels of accuracy.} However, the high accuracy is achieved at the expense of considering around $200 \times$ more features than the baseline (compare last rows of Table~\ref{table:TF-IDF} and Table~\ref{table:histogram}).

Fig.~\ref{fig:encoded} illustrates the accuracy of different ML algorithms corresponding to {character-histogram + endianness} features and character-level (1,2,3)-gram TF-IDF features  when base16, base32, base64, and base85 encoded binaries are used. Our results show that  character-level (1,2,3)-gram TF-IDF features has a higher accuracy than the {character-histogram + endianness} features across all  ML models. Character-level encoded (1,2,3)-gram TF-IDF  features provide the noise reduction and increased generalizability advantages of (1,2,3)-gram TF-IDF features.  It also provides increased number of fine grained features to capture bit patterns specific to architectures that are smaller than 8-bits (e.g., nibble patterns of opcodes). More importantly, using encoded character-level (1,2,3)-gram TF-IDF  features requires only $\sum_{i=1}^{3} 16^i$ features compared to  $\sum_{i=1}^{3} 256^i$ features corresponding to the byte-level (1,2,3)-gram TF-IDF features.
\subsection{Quality of the features via t-SNE}

\begin{figure*}[!h]
\floatsetup{capposition = below, floatrowsep =qquad,}
\centering
\ffigbox{%
\begin{subfloatrow}
\centering
\ffigbox[0.5\textwidth]{\caption{Character-histogram + Endianness features of Base16 binaries}}{%
\includegraphics[width=77mm, height = 50mm, scale=0.5]{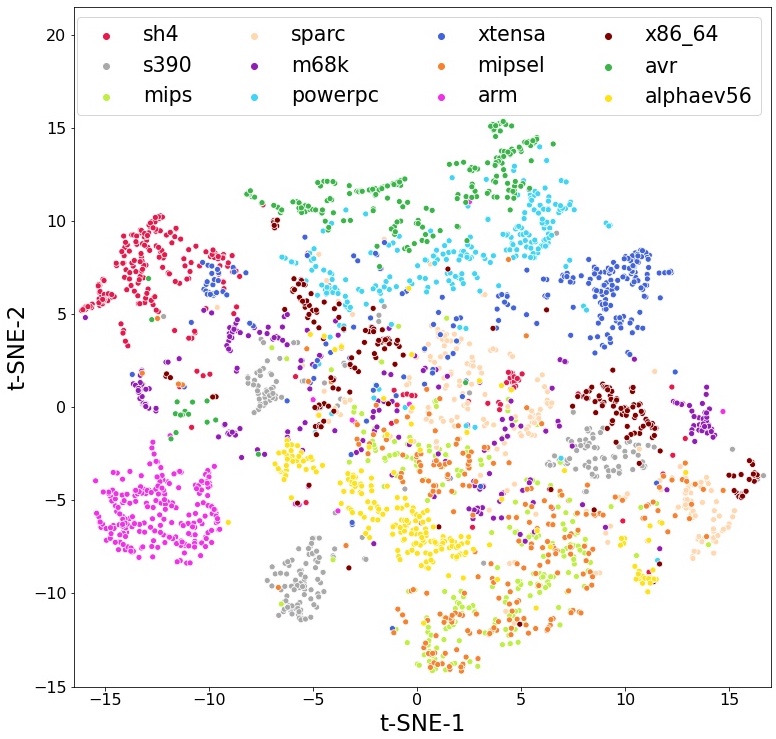}
}
\ffigbox[0.5\textwidth]{\caption{Character-level (1,2,3)-gram TF-IDF features of Base16 binaries}}{%
\includegraphics[width=77mm, height = 50mm, scale=0.5]{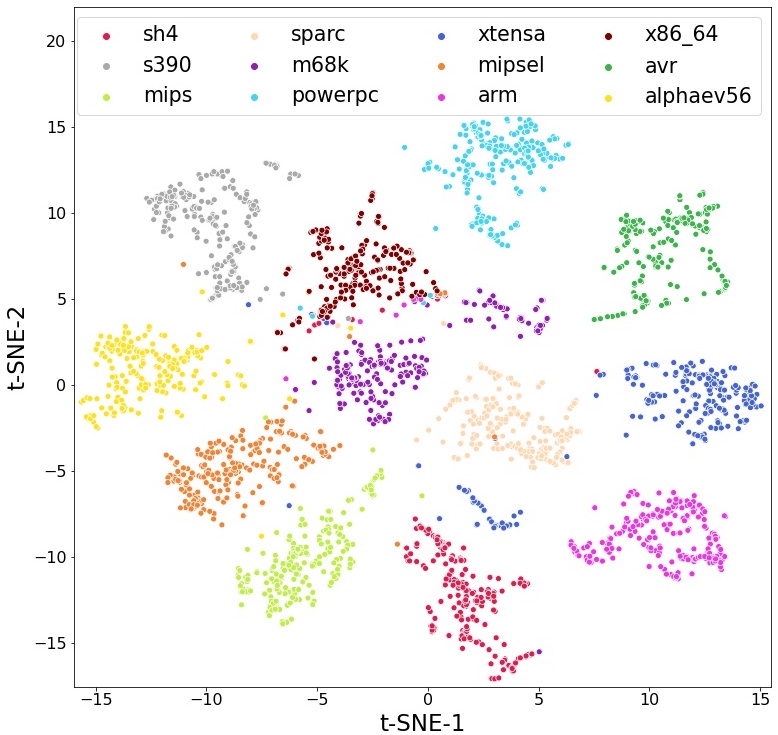}
}
\end{subfloatrow}
\begin{subfloatrow}
\centering
\ffigbox[0.5\textwidth]{\caption{Character-histogram + Endianness features of Base32 binaries}}{%
\includegraphics[width=77mm, height = 50mm, scale=0.5]{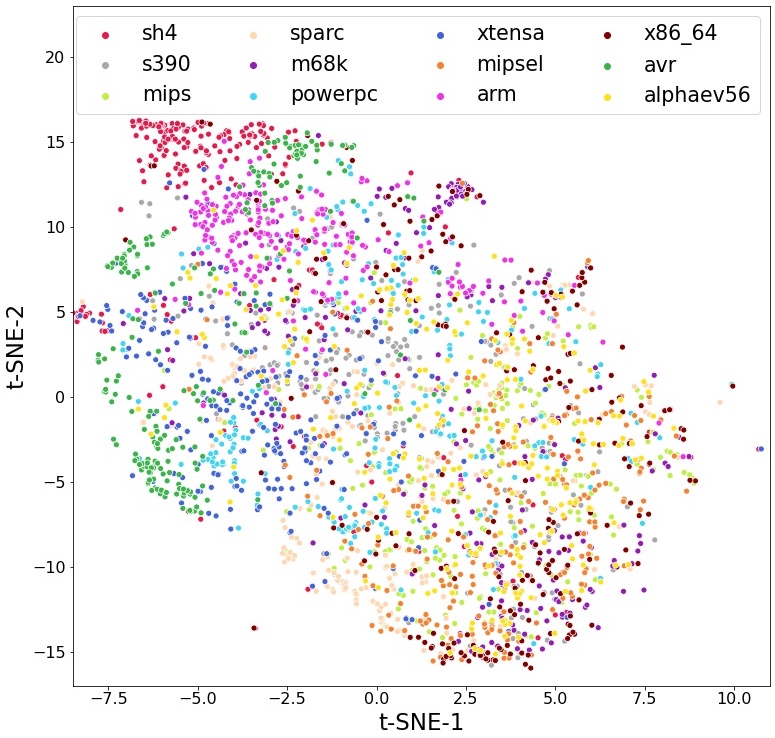}
}
\ffigbox[0.5\textwidth]{\caption{Character-level (1,2,3)-gram TF-IDF features of Base32 binaries}}{%
\includegraphics[width=77mm,height = 50mm, scale=0.5]{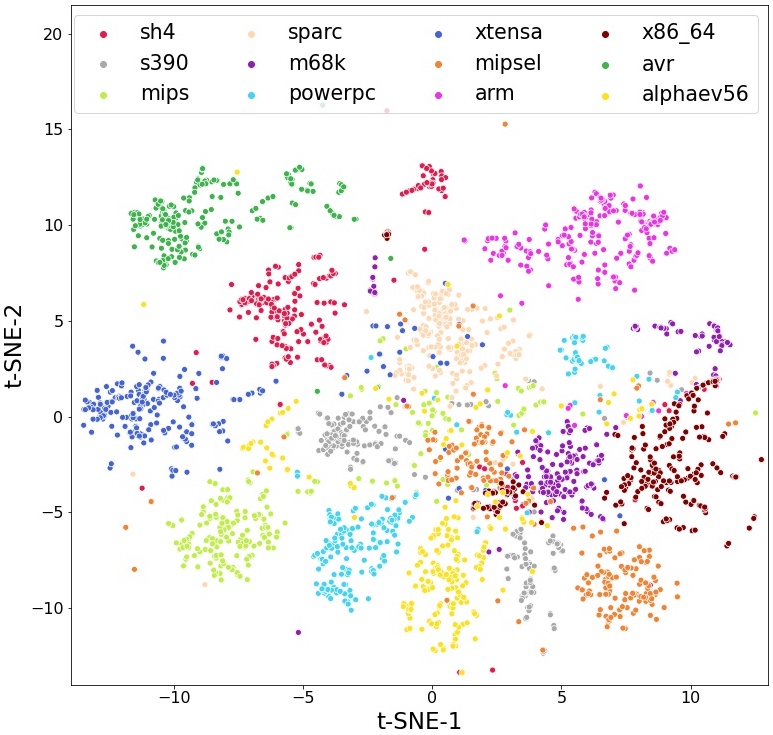}
}
\end{subfloatrow}
\begin{subfloatrow}
\centering
\ffigbox[0.5\textwidth]{\caption{Character-histogram + Endianness features of Base64 binaries}}{%
\includegraphics[width=77mm, height = 50mm, scale=0.5]{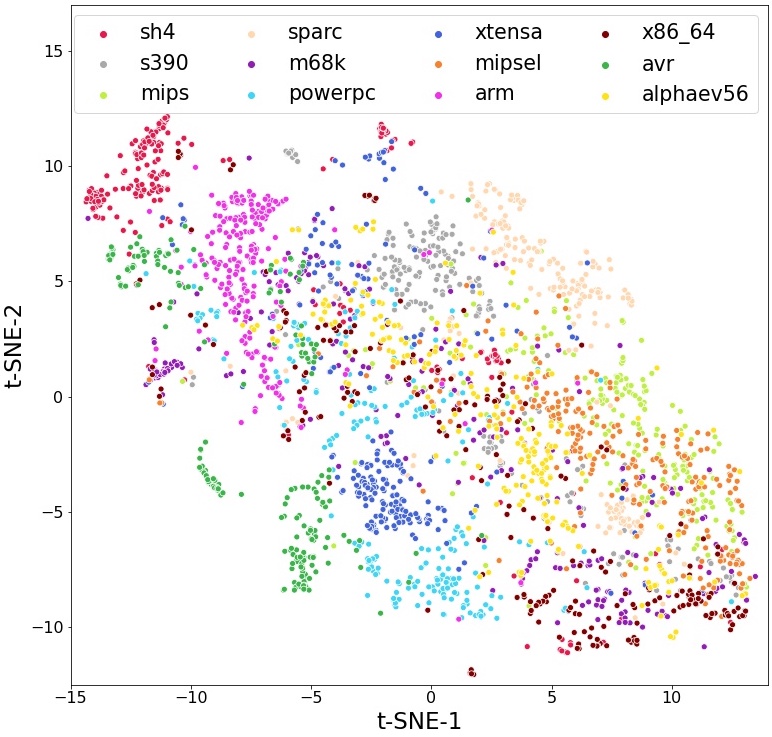}
}
\ffigbox[0.5\textwidth]{\caption{Character-level (1,2,3)-gram TF-IDF features of Base64 binaries}}{%
\includegraphics[width=77mm, height = 50mm, scale=0.5]{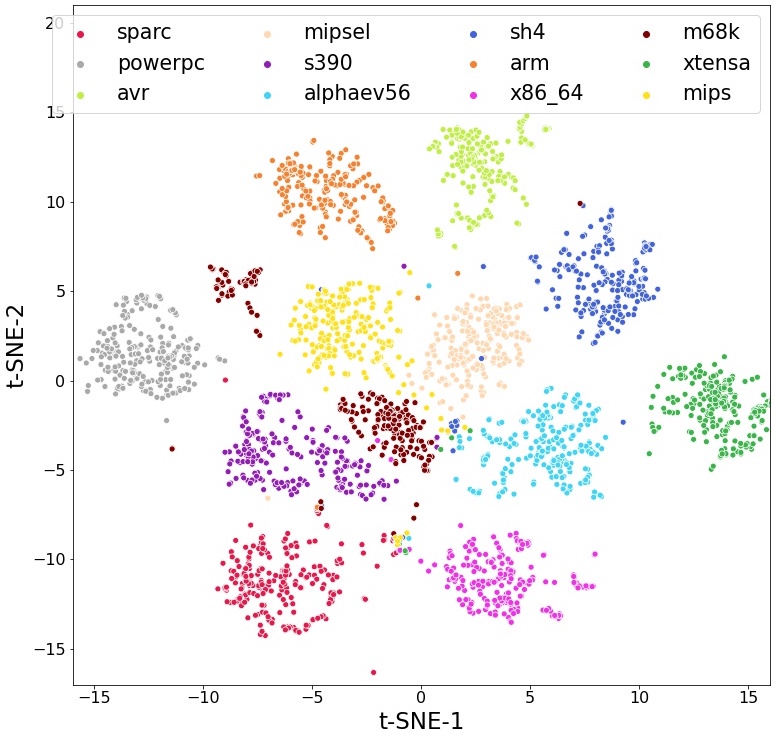}
}
\end{subfloatrow}
\begin{subfloatrow}
\centering
\ffigbox[0.5\textwidth]{\caption{Character-histogram + Endianness features of Base85 binaries}}{%
\includegraphics[width=77mm, height = 50mm, scale=0.5]{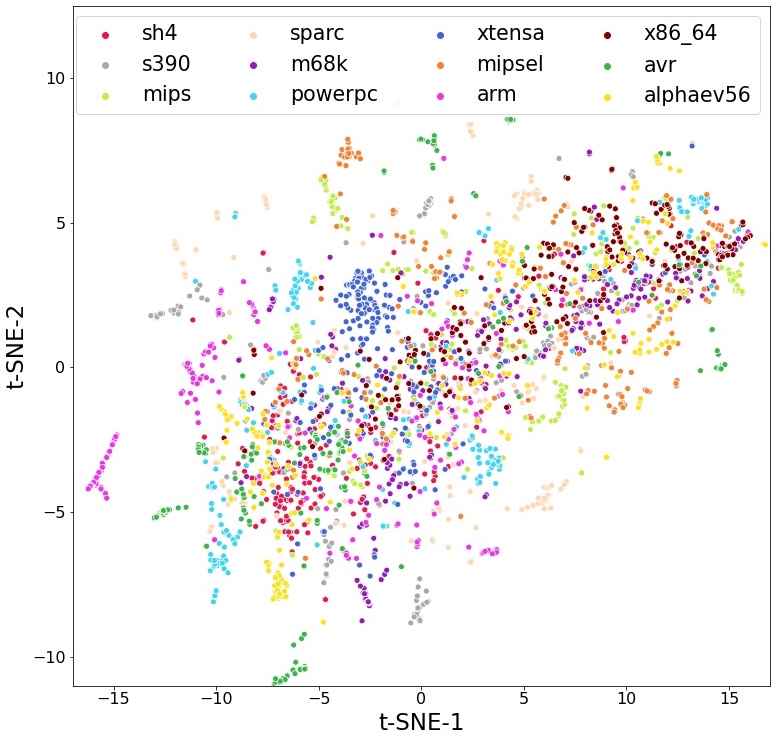}
}
\ffigbox[0.5\textwidth]{\caption{Character-level (1,2,3)-gram TF-IDF features of Base85 binaries}}{%
\includegraphics[width=77mm, height = 50mm, scale=0.5]{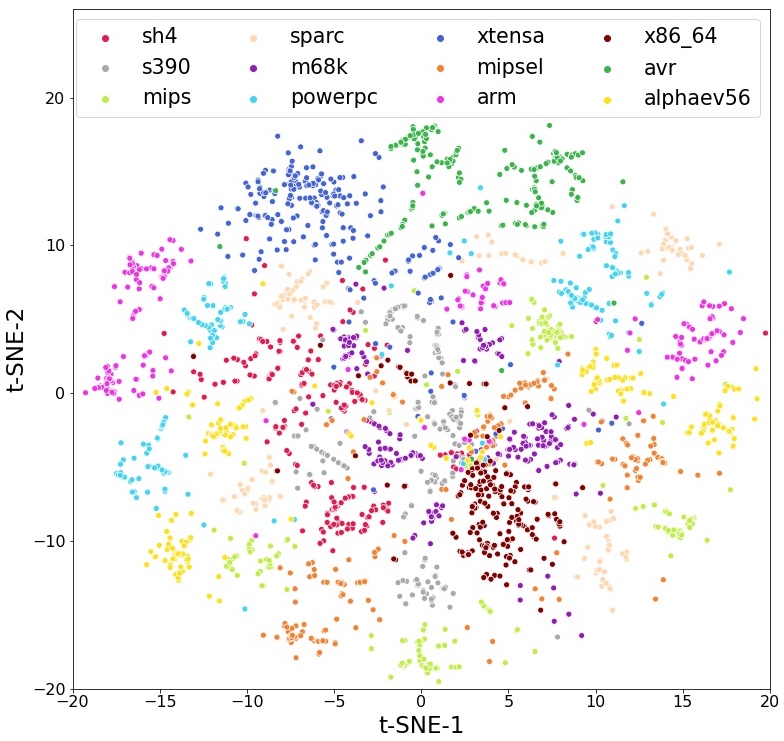}
}
\end{subfloatrow}
}{
\caption{2-D t-SNE plots corresponding to byte-histogram + endianness features and character-level N-gram TF-IDF features extracted from randomly chosen training datasets consist of Base16, Base32, Base64, and Base85 encoded binaries. The character-level (1,2,3)-gram TF-IDF features provide better separability compared to the character-histogram + endianness features.}
\label{fig:tsne-encoded}}
\end{figure*}

Fig.~6 compares the 2-D t-SNE representations of histogram + endianness features and N-gram TF-IDF features of the decoded binaries. Our experiments suggest that byte-level N-gram TF-IDF features result in better separation of the data points corresponding to the different architectures. In contrast, using histogram + endianness features lead to data points of mips and mipsel architectures being indistinguishable. 
This explains the high classification accuracy achieved using the proposed byte-level N-gram TF-IDF features.

Fig.~7 shows the 2-D t-SNE representations of histogram + endianness features and N-gram TF-IDF features of the base16, base32, base64, and base85 encoded binaries. Our experiments show that baseline histogram + endianness features results in poor separation of the clusters related to the data points corresponding to the different architectures. On the other hand, N-gram TF-IDF features extracted from Base16 encoded binaries provide a better separation for the clusters related to different architectures. In fact, comparing with Fig.~6, we can observe this separation is slightly better than the separation achieved via byte-level N-gram TF-IDF features.

\subsection{Number of training data required}

\begin{figure*}[!h]
    \centering
    \includegraphics[width=0.725\textwidth]{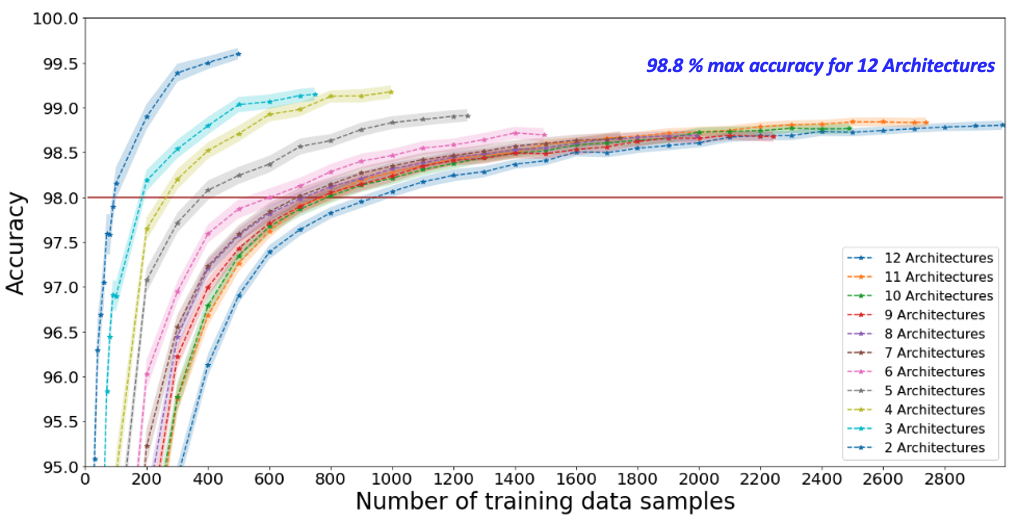}
    \captionof{figure}{SVM accuracy results under different number of architectures when (1,2,3)-gram TF-IDF features are used. The red line indicates the 98\% accuracy margin.}
    \label{fig:SVM-acc-len}
\end{figure*}

\begin{figure*}[!h]
    \centering
    \includegraphics[width=0.725\textwidth]{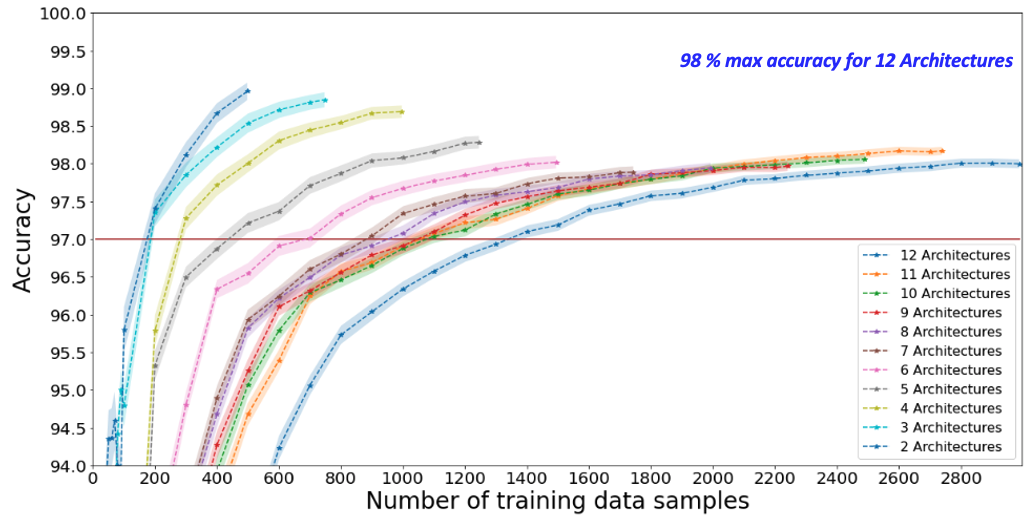}
    \captionof{figure}{LR accuracy results under different number of architectures when (1,2,3)-gram TF-IDF features are used. The red line indicates the 97\% accuracy margin.}
    \label{fig:LR-acc-len}
\end{figure*}

Fig.~\ref{fig:SVM-acc-len} and Fig.~\ref{fig:LR-acc-len} plot accuracy values against the number of training data when classifying different number of ISAs using the  Support Vector Machine (SVM) and Logistic Regression (LR) ML models, respectively. These experiments use byte-level (1,2,3)-gram TF-IDF features extracted from decoded binaries. Our results show that 
byte-level (1,2,3)-gram TF-IDF features achieve high accuracy in both SVM ( $>$ 98 \%) and LR ( $>$ 97\%) consistently under all architecture scenarios. Moreover, in SVM only around 1000 binaries (84 binaries per architecture) are required in the training data set to achieve accuracy $>$ 98 \%. In the case of LR, 1300 binaries (103 binaries per architecture) are required in the training data set to achieve accuracy $>$ 97 \%. This shows that byte-level (1,2,3)-gram TF-IDF features does not require large number of training data to achieve high accuracy. Similar results can be observed in the case of character-level (1,2,3)-gram TF-IDF features extracted from encoded binaries.

\section{Conclusion}\label{sec:conclusion}


In this paper we proposed binary object code feature extraction methods based on $N$-gram Term Frequency-Inverse Document Frequency (TF-IDF) feature model for instruction set architecture (ISA) identification. We used byte-level $N$-gram TF features to extract the successive bytes patterns inherent to architectures. Setting $N=1$, can  recover a class of object code features used in the literature called byte histogram features. However, such approaches require additional domain knowledge/heuristic-based features for capturing successive byte patterns inherent to ISAs and they may be absent in partial binaries.
Histogram-based features are also susceptible to noisy byte values. {\color{black} Hence, histogram and signature-based approaches fail to achieve high accuracy for the binaries with limited data that is corrupted by noise. }
We scaled byte-level $N$-gram TF features by their respective IDF values to attenuate the effect of noisy byte data. Using a 12-architecture dataset, we showed byte-level (1,2,3)-gram TF-IDF features are adequate to achieve high accuracy performance in the Machine Learning (ML)-based ISA identification models. We observed instruction bytes have architecture specific fine grained bit patterns and extracted such patterns using character-level $N$-gram TF-IDF features of encoded binaries (e.g., base16, base32, base64, base85). 
We observed character-level (1,2,3)-gram TF-IDF features of encoded binaries achieving high accuracy while only using less number of features (up to $\approx 16 \times$) compared to the byte-level (1,2,3)-gram TF-IDF features. {\color{black}Our binary code feature extraction methods do not require any prior domain specific knowledge on the ISAs and hence, easily extendable to ISA identification with different number of ISAs.} Promising future research directions include investigation of the effect of NLP and binary-to-text encoding-based object code feature extraction methods in the fields of file type identification and malware binary detection.


\bibliographystyle{IEEEtran}      
\bibliography{TIFS-Binary_references}

\vskip 0pt plus -20fil
\begin{IEEEbiography}[{\includegraphics[width=1in,height=1.25in,clip,keepaspectratio]{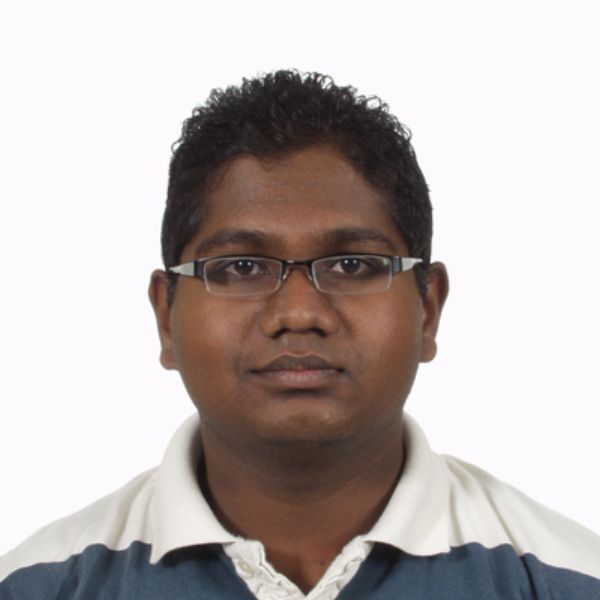}}]{Dinuka Sahabandu }
	is a Ph.D. candidate in the Department of Electrical and Computer Engineering at the University of Washington - Seattle. He received the B.S. degree and M.S. degree in Electrical Engineering from the Washington State University - Pullman in 2013 and 2016, respectively.  His research interests include game theory for network security and control of multi-agent systems.
\end{IEEEbiography}
\vskip 1pt plus -1fil
\begin{IEEEbiography}[{\includegraphics[width=1in,height=1.25in,clip,keepaspectratio]{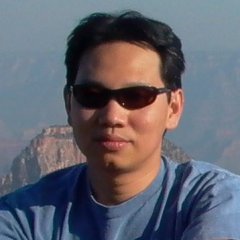}}]{J. Sukarno Mertoguno }
	is a Professor in the School of Cybersecurity and Privacy at Georgia Tech. He previously served as Chief Innovation Officer at GTRI.
He received his Ph.D. from SUNY-Binghamton. His education background includes theoretical physics and electrical engineering.
Before joining Georgia Tech, Dr. Mertoguno managed basic and applied scientific research in cybersecurity and complex software for the Office of Naval Research (ONR), and prior to ONR he was a system and chip architect in the San Francisco Bayt Area.
He developed several novel concepts, such as BFT++, Learn2Reason, CryptoFactory, and bottom-up formal methods. He has led major projects including DARPA PAPPA, DARPA AIMEE, DARPA ReMath, DARPA REPO, DARPA V-SPELLS, and US-Israel Consortium on Cybersecurity for Energy Sector.
\end{IEEEbiography}
\vskip 1pt plus -1fil
\begin{IEEEbiography}[{\includegraphics[width=1in,height=1.25in,clip,keepaspectratio]{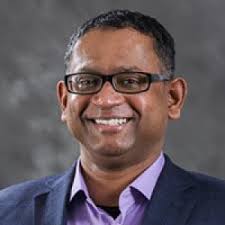}}]{Radha Poovendran}(F'15) is a Professor in the Department of Electrical and Computer Engineering at the University of Washington (UW) - Seattle. He served as the Chair of the Electrical and Computer Engineering Department at UW for five years starting January 2015. He is the Director of the Network Security Lab (NSL) at UW. He is the Associate Director of Research of the UW Center for Excellence in Information Assurance Research and Education. He received the B.S. degree in Electrical Engineering and the M.S. degree in Electrical and Computer Engineering from the Indian Institute of Technology- Bombay and University of Michigan - Ann Arbor in 1988 and 1992, respectively. He received the Ph.D. degree in Electrical and Computer Engineering from the University of Maryland - College Park in 1999. His research interests are in the areas of wireless and sensor network security, control and security of cyber-physical systems, adversarial modeling, smart connected communities, control-security, games-security, and information theoretic security in the context of wireless mobile networks. He is a Fellow of the IEEE for his contributions to security in cyber-physical systems. He is a recipient of the NSA LUCITE Rising Star Award (1999), National Science Foundation CAREER (2001), ARO YIP (2002), ONR YIP (2004), and PECASE (2005) for his research contributions to multi-user wireless security. He is also a recipient of the Outstanding Teaching Award and Outstanding Research Advisor Award from UW EE (2002), Graduate Mentor Award from Office of the Chancellor at University of California - San Diego (2006), and the University of Maryland ECE Distinguished Alumni Award (2016). He was co-author of award-winning papers including IEEE/IFIP William C. Carter Award Paper (2010) and WiOpt Best Paper Award (2012). 
\end{IEEEbiography}

\end{document}